\documentclass[aps,prb,twocolumn]{revtex4-1}

\usepackage{hyperref}
\usepackage[utf8]{inputenc}
\usepackage{amsmath}
\usepackage{bm}
\usepackage{graphicx}
\usepackage{verbatim}
\usepackage{color}

\begin{document}

\title{Determination of scattering time and of valley occupation in transition-metal dichalcogenides doped by field effect}
\date{\today}
\author{Thomas Brumme}
\email{thomas.brumme@mpsd.mpg.de}
\altaffiliation{Current address: Max-Planck-Institut f\"ur Struktur und Dynamik der Materie, Luruper Chaussee 149, 22761 Hamburg, Germany}
\affiliation{CNRS, UMR 7590, Sorbonne Universit\'{e}s, UPMC Univ Paris 06, IMPMC - Institut de Min\'{e}ralogie, de Physique des Mat\'{e}riaux, et de Cosmochimie, 4 place Jussieu, F-75005, Paris, France}
\author{Matteo Calandra}
\affiliation{CNRS, UMR 7590, Sorbonne Universit\'{e}s, UPMC Univ Paris 06, IMPMC - Institut de Min\'{e}ralogie, de Physique des Mat\'{e}riaux, et de Cosmochimie, 4 place Jussieu, F-75005, Paris, France}
\author{Francesco Mauri}
\affiliation{Dipartimento di Fisica, Universit\`{a} di Roma La Sapienza, Piazzale Aldo Moro 5, I-00185 Roma, Italy}

\begin{abstract}
The transition-metal dichalcogenides have attracted a lot of attention as a possible stepping-stone toward atomically
thin and flexible field-effect transistors. One key parameter to describe the charge transport is the time between two
successive scattering events -- the transport scattering time. In a recent report, we have shown that it is possible
to use density functional theory to obtain the band structure of two-dimensional semiconductors in presence of field
effect doping. Here, we report a simple method to extract the scattering time from the experimental conductivity and
from the knowledge of the band structure. We apply our approach to monolayers and multilayers of MoS$_2$, MoSe$_2$, MoTe$_2$,
WS$_2$, and WSe$_2$ in presence of a gate. In WS$_2$, for which accurate measurements of mobility have been published, we find
that the scattering time is inversely proportional to the density of states at the Fermi level. Finally, we show
that it is possible to identify the critical doping at which different valleys start to be occupied from the
doping-dependence of the conductivity.
\end{abstract}


\maketitle

\begin{figure*}
 \includegraphics[scale=0.58,clip=]{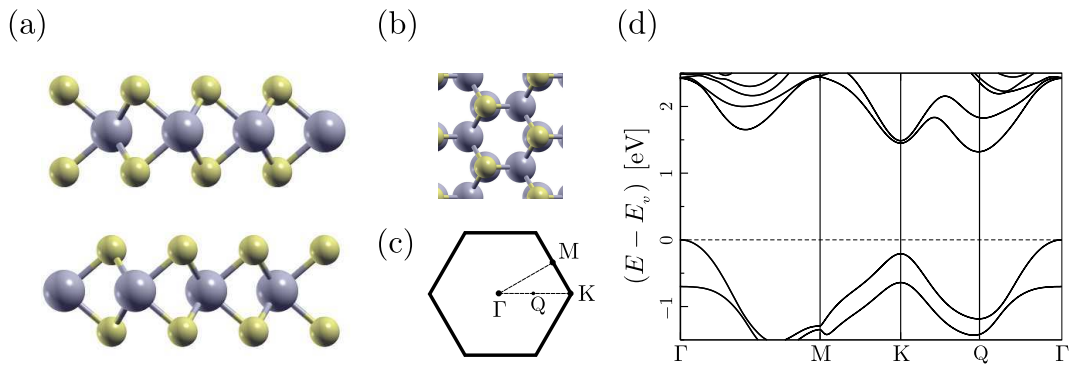}
 \caption{\label{fig:overview}(Color online) In the 2H structure of the TMDs the transition-metal atoms of one layer
          are on top the chalcogen atoms of the other layer while the atoms within a layer form a hexagonal
          pattern as can be seen in the (a) side and (b) top view of the structure (grey -- transition metal, yellow -- chalcogen).
          The hexagonal 2D Brillouin zone shown in (c) has the special points $\Gamma$, M, and K. In the conduction
          band the TMDs have a minimum approximately halfway between $\Gamma$ and K (hereby labeled Q point) as shown in the band structure in (d).
          Its position with respect to the minimum at K depends on e.g. strain and the number of layers.
          The spin-orbit splitting of the bands at K is larger for the TMDs with tungsten
          (in (d) the band structure for bilayer WS$_2$ is shown).}
\end{figure*}

The discovery of graphene\cite{geim2007} has also led to an ever-growing interest in other two-di\-men\-sion\-al (2D) materials
such as monolayers or few-layer systems (nanolayers) of transition-metal dichalcogenides\cite{jariwala2014,ganatra2014} (TMDs).
TMDs form layered structures like graphene in which the different layers are held together by weak van der Waals forces (cf. Fig.~\ref{fig:overview}).
Thus, similar to graphene, one can easily extract single or few layers from the bulk compound
using the mechanical-exfoliation or other experimental techniques.
Different from graphene, TMDs, have an intrinsic band gap (cf. Fig.~\ref{fig:overview}(d)) which makes them especially interesting as atomically thin
field-effect transistors (FETs). Furthermore, one can combine different 2D materials to create van der Waals
heterostructures\cite{geim2013} with interesting properties.

One key parameter to understand the charge transport in TMDs is the time between two
successive scattering events as it determines e.g. the mean-free path and the mobility of the charge carriers.
However, the transport scattering time is not easily accessible by experiments.
Here, we report a simple method to extract the scattering time for field-effect doping of monolayers and
multilayers of MoS$_2$, MoSe$_2$, MoTe$_2$, WS$_2$, and WSe$_2$. More specifically, we calculate the ratio
of the Hall mobility to the scattering time $\mu_\mathrm{Hall}/\tau$ using Boltzmann transport theory within
the relaxation-time approximation. Thus, one can easily extract the scattering time by comparing the measured
Hall mobility with this calculated ratio. We exemplify the extraction for WS$_2$ for measurements done by
Braga et al. in Ref.~\citenum{braga2012}.
We furthermore show that the onset of doping of the different conduction-band minima in TMDs can be determined
by a simple conductivity measurement. This allows for an estimation of the energy difference between different
minima in the undoped compounds (cf. Fig.~\ref{fig:overview}(d)).

In order to extract the transport scattering time $\tau$ from experiments we use the Boltzmann transport equation\cite{madsen2006,ashcroft} (BTE)
to calculate the conductivity tensor and the Hall tensor within the relaxation-time approximation.
The conductivity tensor $\sigma_{\alpha\beta}$ in 2D can be written as
\begin{align}
\label{eq:sigma_full}
\sigma_{\alpha\beta}(T;E_F) &= \frac{e^2}{\left(2\pi\right)^2}\sum_i\int\tau_{i,\mathbf{k}}\,v_\alpha^{i,\mathbf{k}}\,v_\beta^{i,\mathbf{k}}\notag\\
                            &\quad   \times\left[-\frac{\partial f_{E_F}(T;\varepsilon_{i,\mathbf{k}})}{\partial\varepsilon}\right]d\mathbf{k},
\end{align}
where $e$ is the charge of the charge carriers in band $\varepsilon_{i,\mathbf{k}}$ with momentum $\mathbf{k}$,
$f_{E_F}(T;\varepsilon)$ is the Fermi function $f_{E_F}(T;\varepsilon)=\left(\exp\left[\left(\varepsilon-E_F\right)/\left(k_BT\right)\right]+1\right)^{-1}$,
and $v_\alpha^{i,\mathbf{k}}=1/\hbar\:\partial\varepsilon_{i,\mathbf{k}}/\partial k_\alpha$ is the group velocity.
The scattering time $\tau_{i,\mathbf{k}}$ depends on both the band index $i$ and the $\mathbf{k}$-vector direction.
For crystals with hexagonal symmetry such as TMDs the conductivity has only two independent coefficients\cite{ashcroft}
(in-plane $\sigma_{||}\equiv\sigma$ and out-of-plane $\sigma_{zz}$ component).
The (in-plane) Hall mobility, which is measured in the experiments, can be written as
\begin{align}
 \mu_\mathrm{Hall}(T;E_F)&=\sigma(T;E_F)\,R_{xyz}(T;E_F),
\end{align}
where $R_{xyz}(T;E_F)$ is the Hall coefficient for the induced electric field
along $y$ if the current and the magnetic field are applied along $x$ and $z$, respectively.\cite{madsen2006}.
Note that $n_\mathrm{Hall}\equiv R^{-1}_{xyz}(T;E_F)$ is not the same as the
doping charge $n$ and can deviate substantially from the case of an isolated parabolic band. This is due to
the specific band structure of the TMDs as shown in Ref.~\citenum{brumme2015}.
In the following we drop the ``$(T;E_F)$''.

\begin{figure}
 \includegraphics[scale=0.5,clip=]{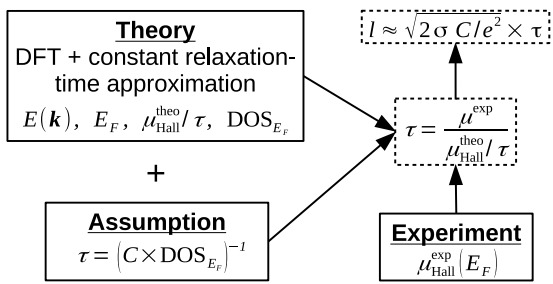}
 \caption{\label{fig:flowchart} Flow chart illustrating the procedure used in this paper to calculate
          the transport properties. First, the band structure is calculated within density-functional
          theory (DFT) for a doping concentration $n$ as has been implemented in Quantum ESPRESSO\cite{quantumespresso,brumme2014}.
          The band structure is used as input for BoltzTraP\cite{madsen2006}
          to calculate the transport properties within the constant-scattering-time approximation using the expansion coefficients.
          Those results can then be used either to extract $\tau$ as shown in Fig.~\ref{fig:mu_ratio_fit} or (assuming $\tau=\left(C\times\mathrm{DOS}_{E_F}\right)^{-1}$)
          to fit to the experimental mobility $\mu_\mathrm{Hall}^\mathrm{exp}$ as shown in Fig.~\ref{fig:mu_full_fit}.}
\end{figure}
\begin{figure}
 \includegraphics[scale=0.38,clip=]{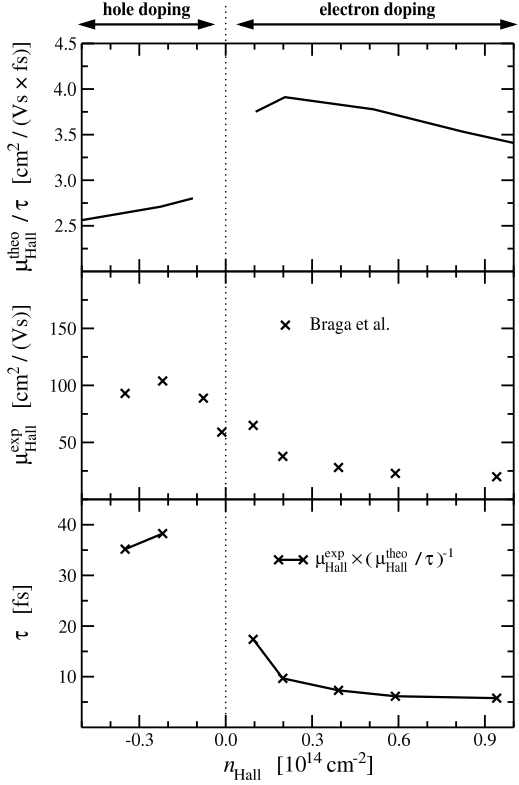}
 \caption{\label{fig:mu_ratio_fit} Extracting the scattering rate for multilayer WS$_2$.
  The top panel shows the calculated ratio $\mu_\mathrm{Hall}^\mathrm{theo}/\tau$ while the measured mobility\cite{braga2012} 
  is given in the middle panel ($T=300\,\mathrm{K}$). The bottom panel shows the ratio of both. As the mobility
  was calculated for doping charge concentrations per unit cell $n$, which in general are different from the measured
  concentration $n_\mathrm{Hall}$\cite{brumme2015}, we linearly interpolated the theoretical mobility
  to extract its value at the experimental doping.}
\end{figure}

For scattering independent of the band index and the wave-vector direction $\tau_{i,\mathbf{k}}=\tau$.
Under this assumption, $\sigma/\tau$, $\mu/\tau$, and $R_{xyz}$ are all independent on $\tau$.
We can then use the ab-initio band structure to
calculate the ratio of the theoretical conductivity $\sigma^\mathrm{theo}$ and Hall mobility $\mu_\mathrm{Hall}^\mathrm{theo}$ to
the scattering time $\tau$.
Our method to calculate those quantities is further clarified in Fig.~\ref{fig:flowchart} and the computational details can be found in the supplemental material.\cite{supmat}
The ratio of the measured Hall mobility $\mu_\mathrm{Hall}^\mathrm{exp}$ to the calculated
ratio $\mu_\mathrm{Hall}^\mathrm{theo}/\tau$ can be used to extract the scattering time from the experimental data.
Figure \ref{fig:mu_ratio_fit} exemplifies this extraction for WS$_2$. The experimental data
of Braga et al. was measured on a thick sample at $T=300\,\mathrm{K}$.
We compare this data with the calculations for trilayer WS$_2$ which is a good approximation for nanolayers with
more than 3 layers and doping larger than $n>10^{13}\,\mathrm{cm}^{-2}$ as shown in Ref.~\citenum{brumme2015}.
The scattering time thus extracted decreases with increasing doping-charge concentration $n_\mathrm{Hall}$ as shown in
the bottom panel of Fig.~\ref{fig:mu_ratio_fit}. Furthermore, electrons are scattered more frequently than holes.
Hole and electron mobilities are still comparable due to the much higher effective mass of holes at the $\Gamma$ point
as compared to the mass of electrons in the conduction band.\cite{yun2012}

The assumption of constant scattering time is not a far-fetched simplification.
In fact, for constant scattering matrix elements
one can show that the scattering rate $1/\tau$ is proportional to the density of states (DOS) at the Fermi energy.
This is also true for electron-phonon interaction if the phonon energy is negligible with respect to the Fermi energy.
Accordingly, because of Matthiessen's rule, the total scattering time is directly proportional to the inverse of
the total DOS at the Fermi energy $\tau=\left(C\times\mathrm{DOS}_{E_F}\right)^{-1}$.

\begin{figure}
 \includegraphics[scale=0.38,clip=]{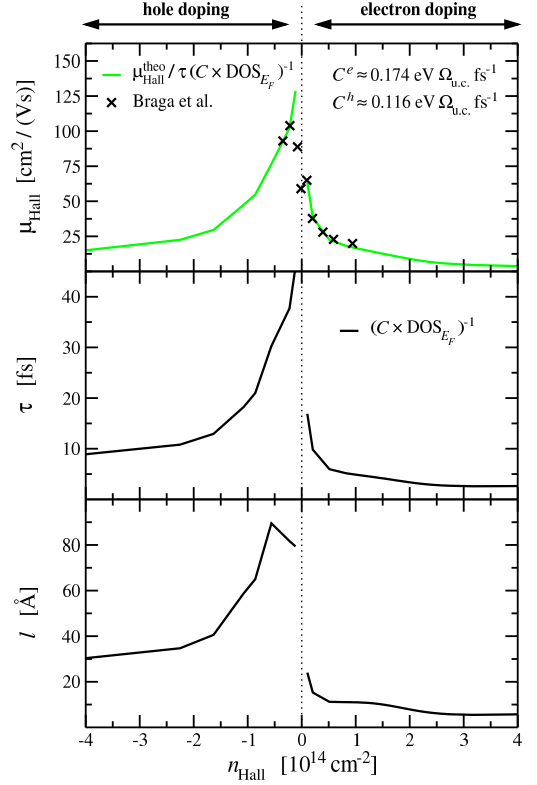}
 \caption{\label{fig:mu_full_fit} Extracting the scattering time for multilayer WS$_2$ at $T=300\,\mathrm{K}$
          using the constant-scattering-time approximation. The experimental data has been taken from Ref.~\citenum{braga2012}.
          Assuming $\tau=\left(C\times\mathrm{DOS}_{E_F}\right)^{-1}$ one can fit the calculated mobility to the measurements (top panel).
          The resulting scattering time is shown in the middle panel. The corresponding mean-free path,
          $l\approx\sqrt{2\langle v_x^2\rangle}\times\tau=\sqrt{2\sigma\,C/e^2}\times\tau$,
          is given in the bottom panel.}
\end{figure}

In order to obtain the constant $C$, we use the calculated DOS as outline in Fig.~\ref{fig:flowchart} and fit $\mu_\mathrm{Hall}^\mathrm{theo}/\tau\,\left(C\times\mathrm{DOS}_{E_F}\right)^{-1}$
to the data of Ref.~\citenum{braga2012} as shown in top panel of Fig.~\ref{fig:mu_full_fit}.
For electron doping and hole doping we find $C^e\approx0.174\,\mathrm{eV}\,\Omega_\mathrm{u.c.}\,\mathrm{fs}^{-1}$ and
$C^h\approx0.116\,\mathrm{eV}\,\Omega_\mathrm{u.c.}\,\mathrm{fs}^{-1}$, respectively, with $\Omega_\mathrm{u.c.}$ the area
of one unit cell.
The resulting scattering time $\tau$ is given in the middle panel.
It decreases with increasing doping and saturates for high electron
doping at $\tau\approx3\,\mathrm{fs}$. Furthermore, in the hole doping case the carriers are less often scattered.
The good qualitative agreement between the experimental and the theoretical mobility shows that an energy- and
momentum-independent scattering time which is proportional to $\mathrm{DOS}_{E_F}$ can be used to estimate the transport scattering
time for doping larger than $n>10^{13}\,\mathrm{cm}^{-2}$. This indicates that the scattering matrix elements are well approximated by
a constant that is independent from the energy and the momentum of the electrons at the Fermi level. Both, scattering
from neutral defects that do not induce mid-gap states and electron-phonon
scattering from non-polar phonons, can be modeled with constant matrix elements and could thus explain our findings.
Finally, it is important to remark that in calculations the relative position of the mimina in the Q and K valleys,
their effective masses, and the density of states of the conduction band do depend both on the functional and the
lattice parameters used in the calculation (see, e.g., Refs.~\citenum{peelaers2012,yun2012,shi2013}). The good agreement between the experimental and our calculated
mobility in Fig.~\ref{fig:mu_full_fit} validates our electronic structure in the doping region $0.1\times10^{14}<n_\mathrm{Hall}<1\times10^{14}$.

\begin{figure*}
 \includegraphics[scale=0.4,clip=]{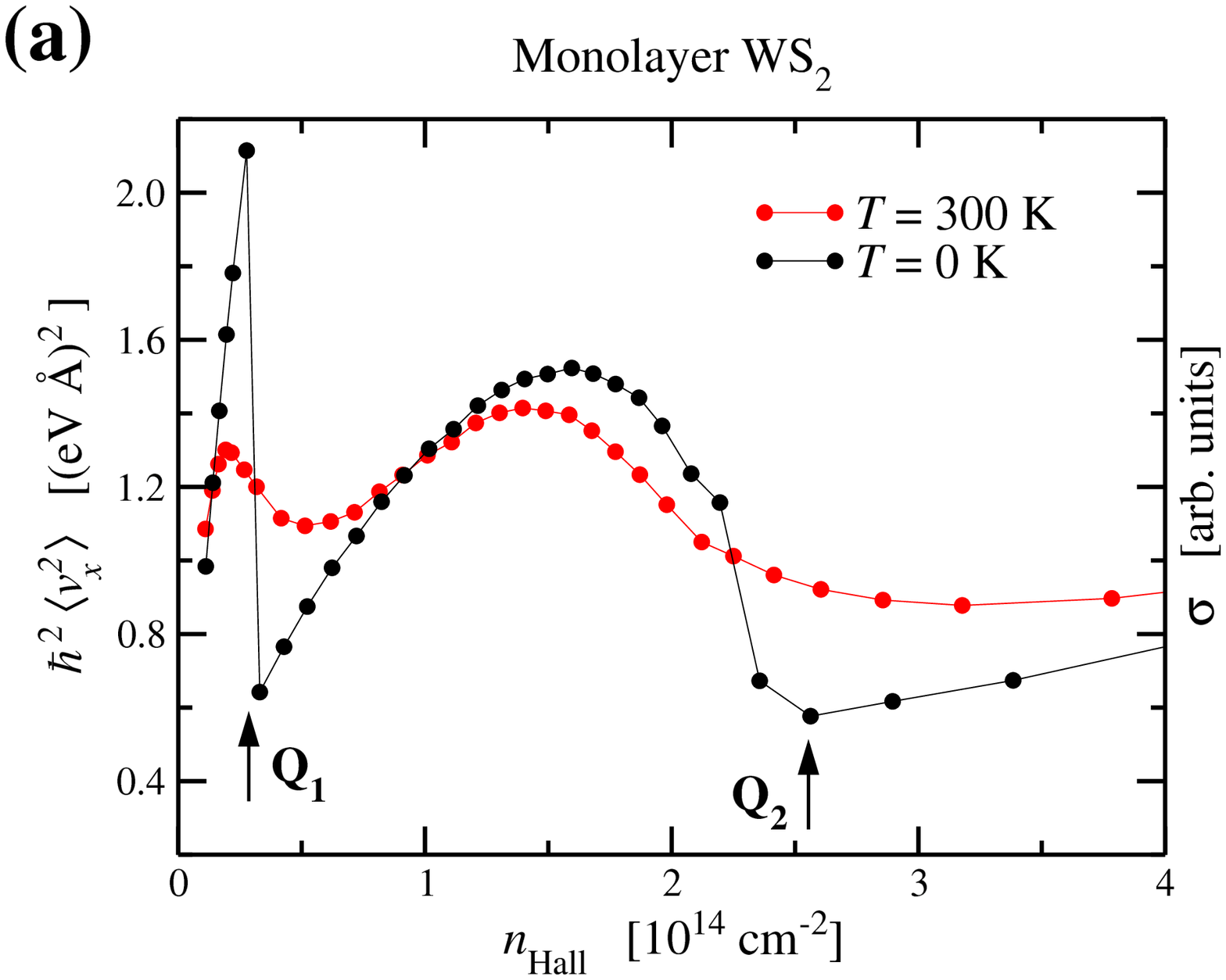}\hspace*{0.5cm}
 \includegraphics[scale=0.4,clip=]{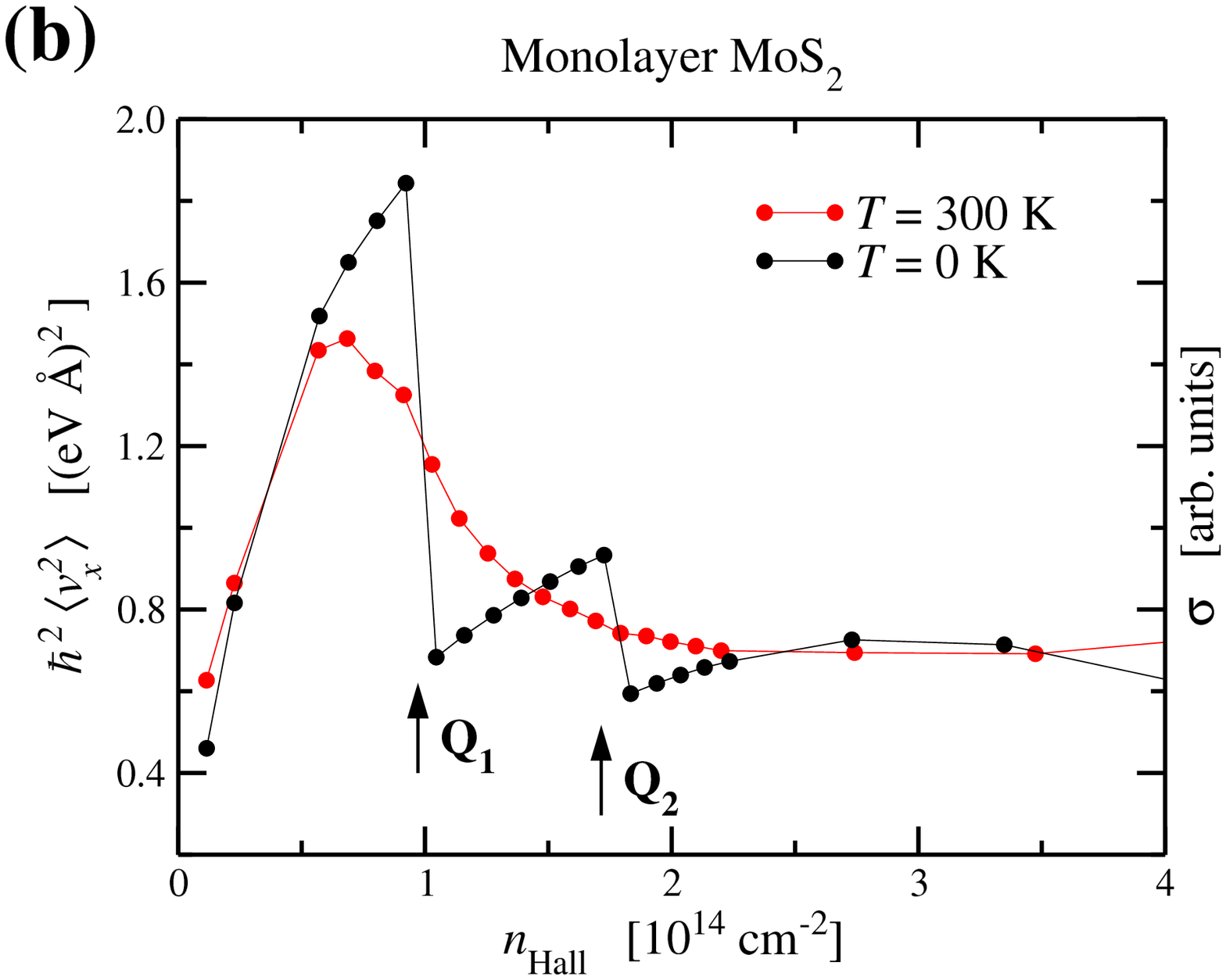}\vspace*{0.5cm}
 \includegraphics[scale=0.4,clip=]{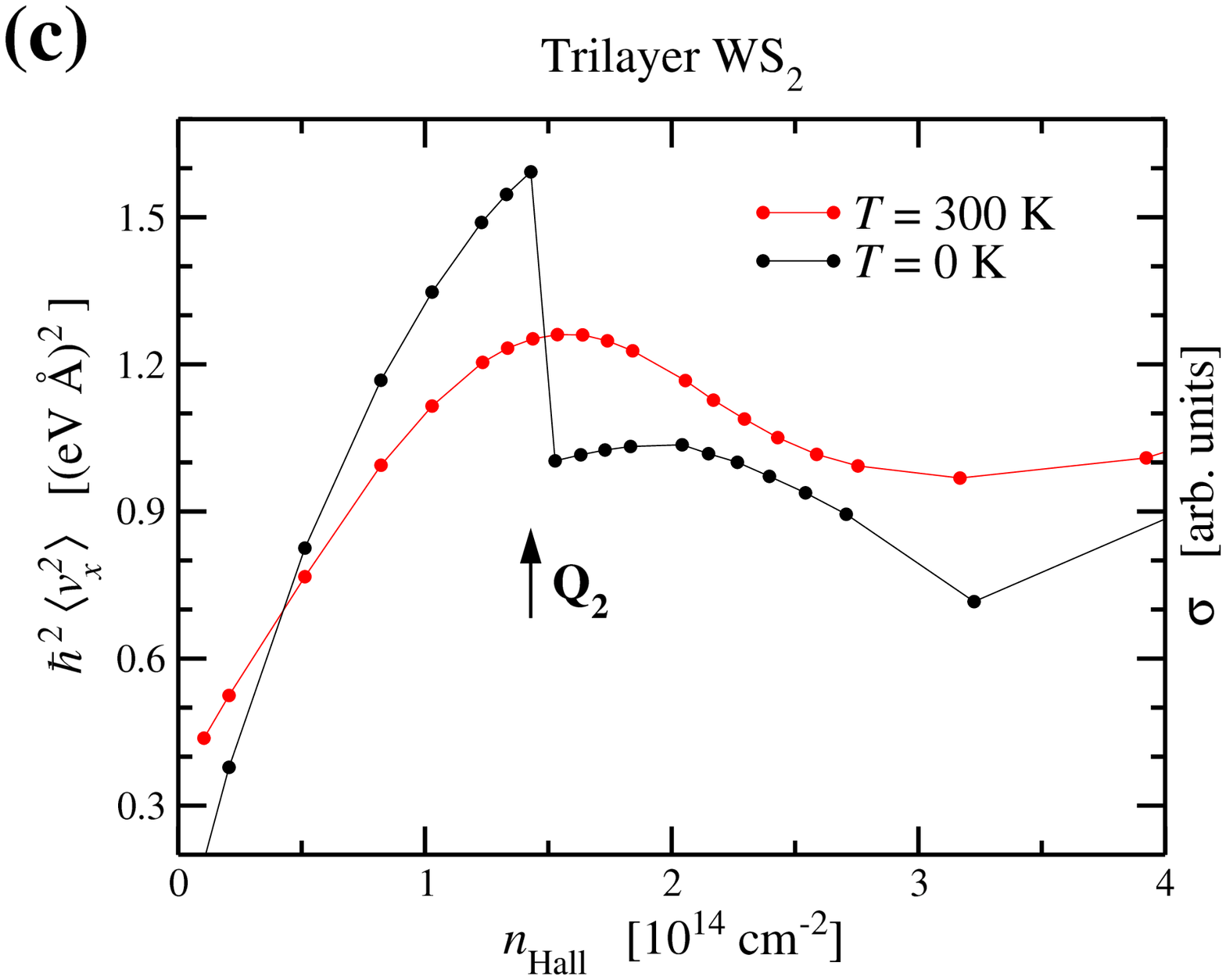}\hspace*{0.5cm}
 \includegraphics[scale=0.4,clip=]{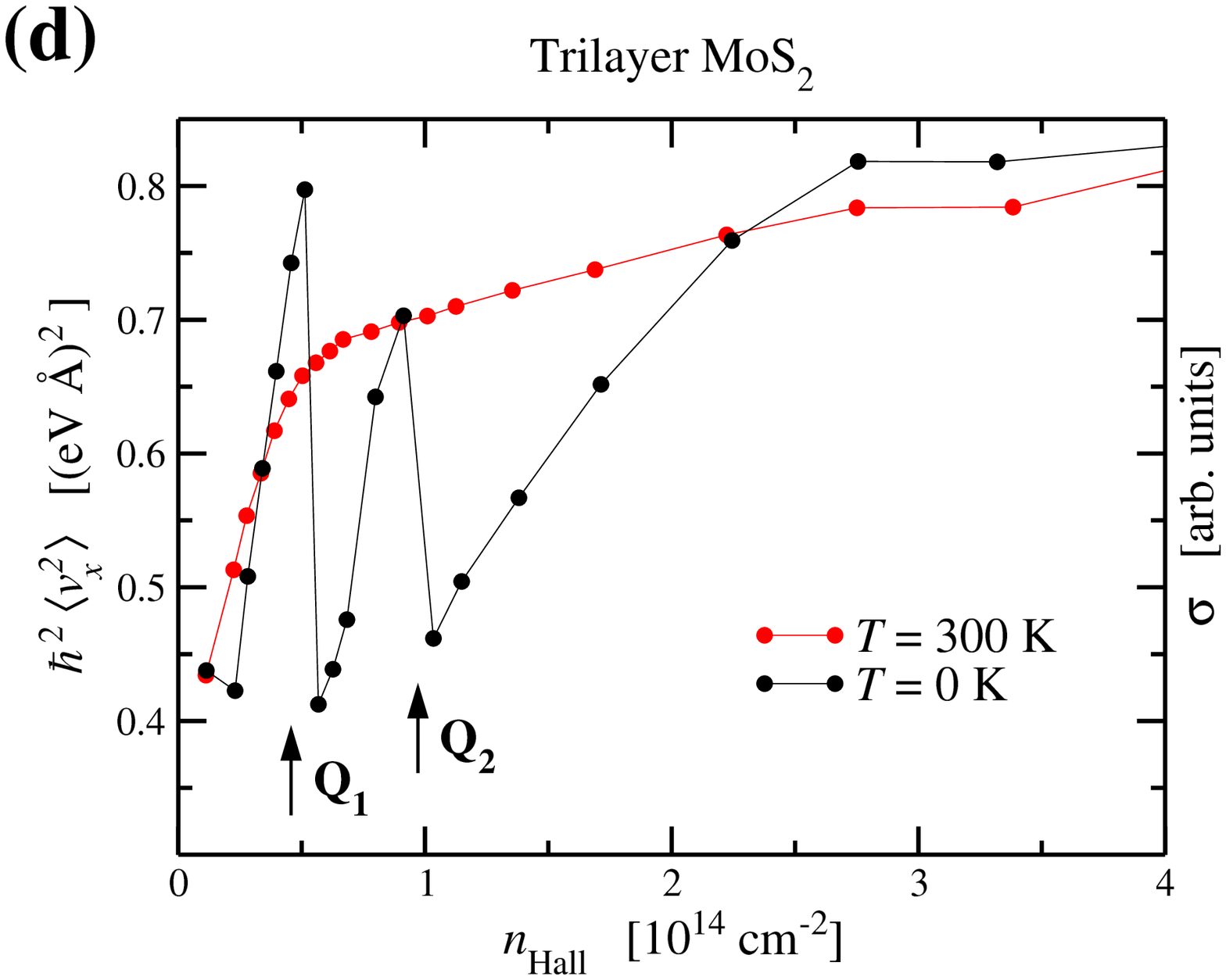}
 \caption{\label{fig:velocity} Average squared velocities $\hbar^2\langle v_x^2\rangle$ of (a,c) WS$_2$
          and (b,d) MoS$_2$ for different number of layers and increasing electron doping. The onset of doping of the different
          bands at the Q point (approximately halfway between $\Gamma$ and K, cf. Fig.~\ref{fig:overview}) for $T=0\,\mathrm{K}$ is indicated by small arrows. For
          trilayer WS$_2$ the doping charge first occupies the band at Q and only the onset of doping of the
          second band at Q is indicated.}
\end{figure*}
Assuming that $\tau=(C\times\mathrm{DOS}_{E_F})^{-1}$, the conductivity is proportional to the average squared velocity of the charge carriers
which allows us to estimate the mean-free path. To show this we have to rewrite the in-plane conductivity as follows:
\begin{align}
\sigma &= e^2\,\tau\,\langle v_x^2\rangle\,\frac{1}{\left(2\pi\right)^2}\sum_i\int\left[-\frac{\partial f_{E_F}(T;\varepsilon_{i,\mathbf{k}})}{\partial\varepsilon}\right]\text{d}^2\mathbf{k}\\
&= e^2\,\tau\,\langle v_x^2\rangle\,\mathrm{DOS}_{E_F} = e^2\,\langle v_x^2\rangle\,C^{-1},
\end{align}
where $\langle v_x^2\rangle$ is the average of the squared in-plane velocity over the Fermi surface
\begin{align}
\label{eq:vsquared}
\langle v_x^2\rangle&=\frac{\sum_i\int\left(v_x^{i,\mathbf{k}}\right)^2\left[-\frac{\partial f_{E_F}(T;\varepsilon_{i,\mathbf{k}})}{\partial\varepsilon}\right]\text{d}^2\mathbf{k}}
                         {\sum_i\int\left[-\frac{\partial f_{E_F}(T;\varepsilon_{i,\mathbf{k}})}{\partial\varepsilon}\right]\text{d}^2\mathbf{k}},
\end{align}
with $v_x^{i,\mathbf{k}}=1/\hbar\:\partial\varepsilon_{i,\mathbf{k}}/\partial k_x$. We furthermore assumed that $\mathrm{DOS}_{E_F}$ is given per unit-cell area.
Note that it is only possible to define the average squared (in-plane) velocity in such a way due to the hexagonal symmetry.
Similarly, using the same approximations, the mobility can be written as $\mu_\mathrm{Hall}\propto\langle v_x^2\rangle/n_\mathrm{Hall}$.
The mean-free path is thus given by $l=\langle v\rangle\,\tau\approx\sqrt{\langle v_x^2+v_y^2\rangle}\times\tau=\sqrt{2\langle v_x^2\rangle}\times\tau=\sqrt{2\sigma\,C/e^2}\times\tau$.
The bottom panel of Fig.~\ref{fig:mu_full_fit} shows $l$ for the sample of Braga et al.\cite{braga2012}.
Since the scattering time for electrons is shorter than for holes the corresponding mean-free path
is also much smaller. Holes can travel several unit cells until they are scattered even for doping
larger than $n>1\times10^{14}\,\mathrm{cm}^{-2}$. The mean-free path for electrons on the other hand
already becomes as small as three unit cells for $n\approx0.25\times10^{14}\,\mathrm{cm}^{-2}$
and even smaller than two unit cell for $n>2.4\times10^{14}\,\mathrm{cm}^{-2}$. This short
mean-free path demonstrates that we in principle need to go beyond the semiclassical BTE to include
quantum effects, which is either an intrinsic problem of WS$_2$ or just a property of this sample due to
a large number of defects. Calculations using more advanced methods such as the nonequilibrium Green's
function method are however beyond the scope of this paper and we thus leave this interesting problem
for future investigations. The assumption of a constant scattering time which is proportional to
$\mathrm{DOS}_{E_F}$ still leads to a good qualitative agreement for $|n|\leq10^{14}\,\mathrm{cm}^{-2}$
as can been seen in the top panel of Fig.~\ref{fig:mu_full_fit}.

In Fig.~\ref{fig:velocity} we show the average squared velocities $\hbar^2\langle v_x^2\rangle$ (i.e. the conductivity assuming $\tau=(C\times\mathrm{DOS}_{E_F})^{-1}$) for electron doping of
the monolayer and trilayer systems of WS$_2$ and MoS$_2$. Corresponding curves for hole doping and MoSe$_2$,
MoTe$_2$, and WSe$_2$ can be found in the supplemental material\cite{supmat}.
In general, the average velocity first increases and then saturates for higher doping concentrations.
Most interestingly, all systems show at least one kink in the average velocity for $T=0\,\mathrm{K}$.
For higher temperatures these kinks are smeared out until they eventually disappear as can be seen in
the case of trilayer MoS$_2$.

In order to understand the origin of this drops in the conductivity we use a simplified 2D
model with isotropic, quadratic dispersion $\varepsilon_{i,k}=\hbar^2 k^2/\left(2m_i\right)+E^0_i$ with
$k^2=k^2_x+k^2_y$ and $E^0_i$ being the bottom of the band $i$. In the zero-temperature limit, the
average squared velocity $\langle v_x^2\rangle=\langle v_y^2\rangle$ can be written as
\begin{align}
\label{eq:v_model}
 \langle v_x^2\rangle &= \frac{\sum_{i,\mathrm{occ}}g_i^v\left(E_F-E^0_i\right)}{\sum_{i,\mathrm{occ}}g_i^v m_i},
\end{align}
with the sums running only over occupied bands $i$ with mass $m_i$, valley degeneracy $g_i^v$ and $E_F$ being the Fermi energy.

\begin{figure}
 \includegraphics[scale=0.4,clip=]{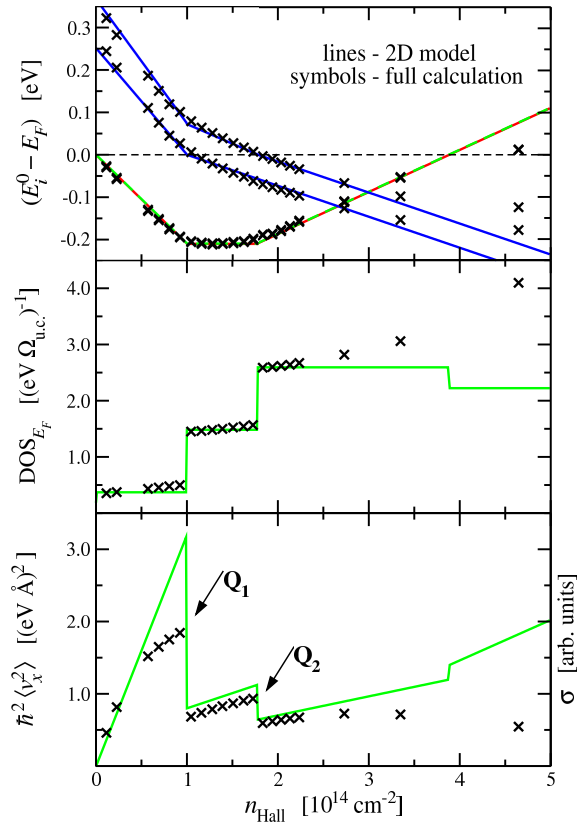}
 \caption{\label{fig:compare_model_calc}(Color online) Comparison between the full ab-initio calculations using the 
          BTE for monolayer MoS$_2$ and the simplified model assuming valleys with isotropic, quadratic dispersion.
          The top panel shows the position of the ab-initio band maxima as symbols. The positions in the
          simplified model are shown as solid lines (blue -- conduction-band minima at Q, green/red -- conduction-band
          minimum at K, the different shadings indicate the 3 different phases mentioned in the text).
          In the middle panel the calculated $\mathrm{DOS}_{E_F}$ is compared with the one of the model
          ($m_\mathrm{K}=0.5\,m_0$ and $m_\mathrm{Q}=m_0$).
          The bottom panel shows the average squared velocity $\hbar^2\langle v_x^2\rangle$ calculated with BoltzTraP
          as symbols and the one obtained with the model as solid lines.
          The onset of doping of the spin-orbit-split bands at the Q point for $T=0\,\mathrm{K}$ is indicated by small arrows.}
\end{figure}
In Fig.~\ref{fig:compare_model_calc} we compare the simplified 2D model with the full ab-initio calculation for a monolayer of MoS$_2$.
The effective masses in the model of $m_\mathrm{K}=0.5\,m_0$ and $m_\mathrm{Q}=m_0$ for electrons at K and at Q, respectively,
were determined using $\mathrm{DOS}_{E_F}$ of the DFT calculations.
The variation of $E^0_i$ as a function of $n_\mathrm{Hall}$ has been adjusted to the ab-initio
calculations by assuming three separate phases as indicated by different shading in the top panel of Fig.~\ref{fig:compare_model_calc}:
(i) first only the two spin-orbit split conduction bands at K are filled until (ii) the first band
at Q touches the Fermi level and the occupation of the K valleys stays constant. Finally, when the second band at Q starts to get filled, the
doping of the K valleys is reduced (iii).
For small doping the model agrees well with the ab-initio results and only for higher doping it starts to deviate.
The main error is the assumption of 2D quadratic bands
with doping-independent mass which leads to an overestimation of the average velocity. Using the effective mass
as calculated by Yun et al.\cite{yun2012} ($m_\mathrm{K}=0.483\,m_0$ and $m_\mathrm{Q}=0.569\,m_0$) the deviation between
model and ab-initio calculations would be even larger.
Note that including a broadening due to a finite temperature only leads to a smooth decrease of $\hbar^2\langle v_x^2\rangle$ instead of a
step close to $n_\mathrm{Hall}=10^{14}\,\mathrm{cm}^{-2}$ but does not decrease the velocity considerably as can be seen in Fig.~\ref{fig:velocity}(b).
The drop is related to the onset of doping of the Q valley. As apparent in eq.~\ref{eq:v_model} it is directly proportional
to $\mathrm{DOS}_{E_F}$. For doping $n_\mathrm{Hall}>2.5\times10^{14}\,\mathrm{cm}^{-2}$ the model shows an increasing average
velocity while the velocity is constant in the ab-initio calculations. These different behaviors are due to the strong
non-parabolicity of the conduction bands at Q.

\begin{figure*}
 \includegraphics[scale=0.4,clip=]{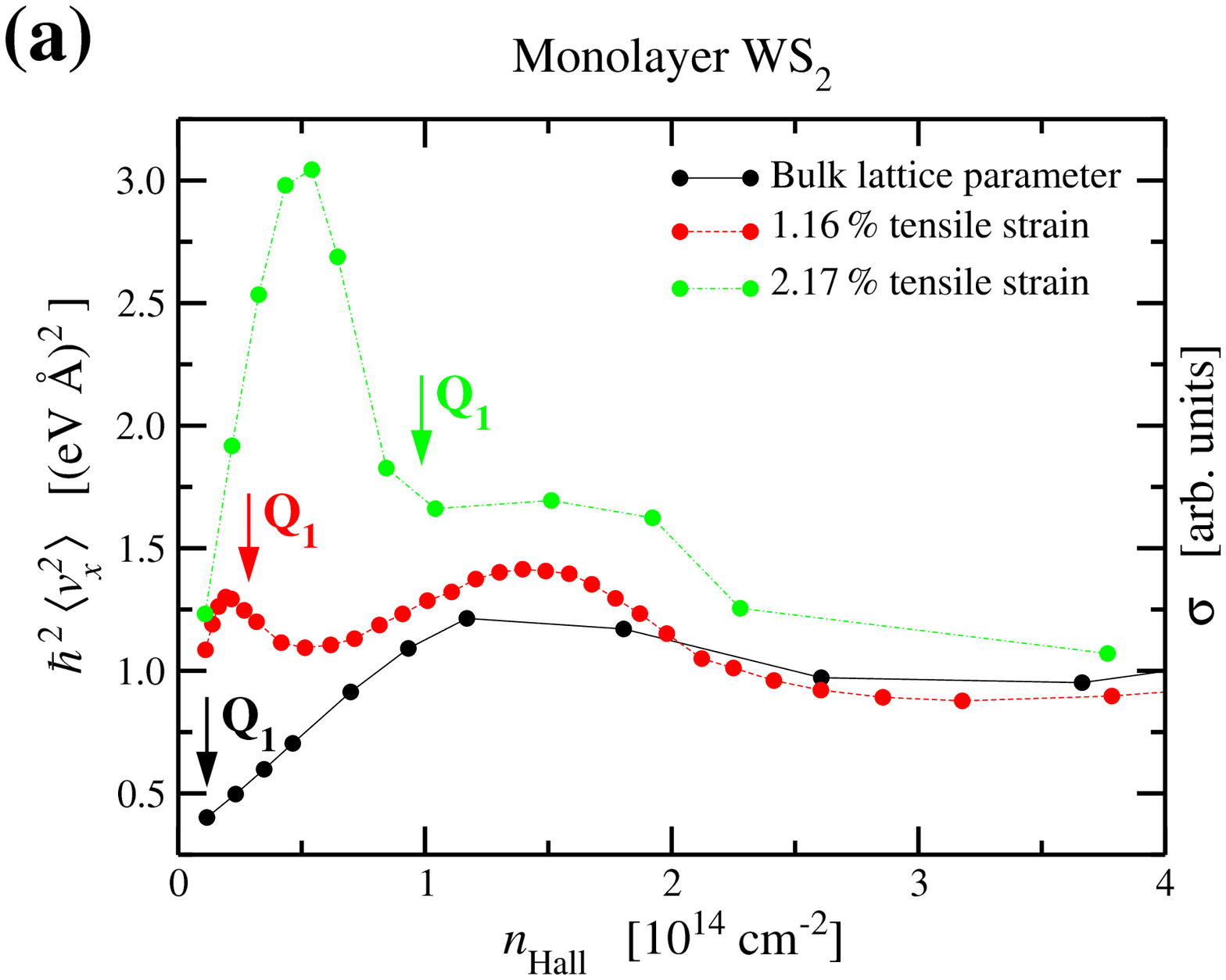}\hspace*{0.5cm}
 \includegraphics[scale=0.4,clip=]{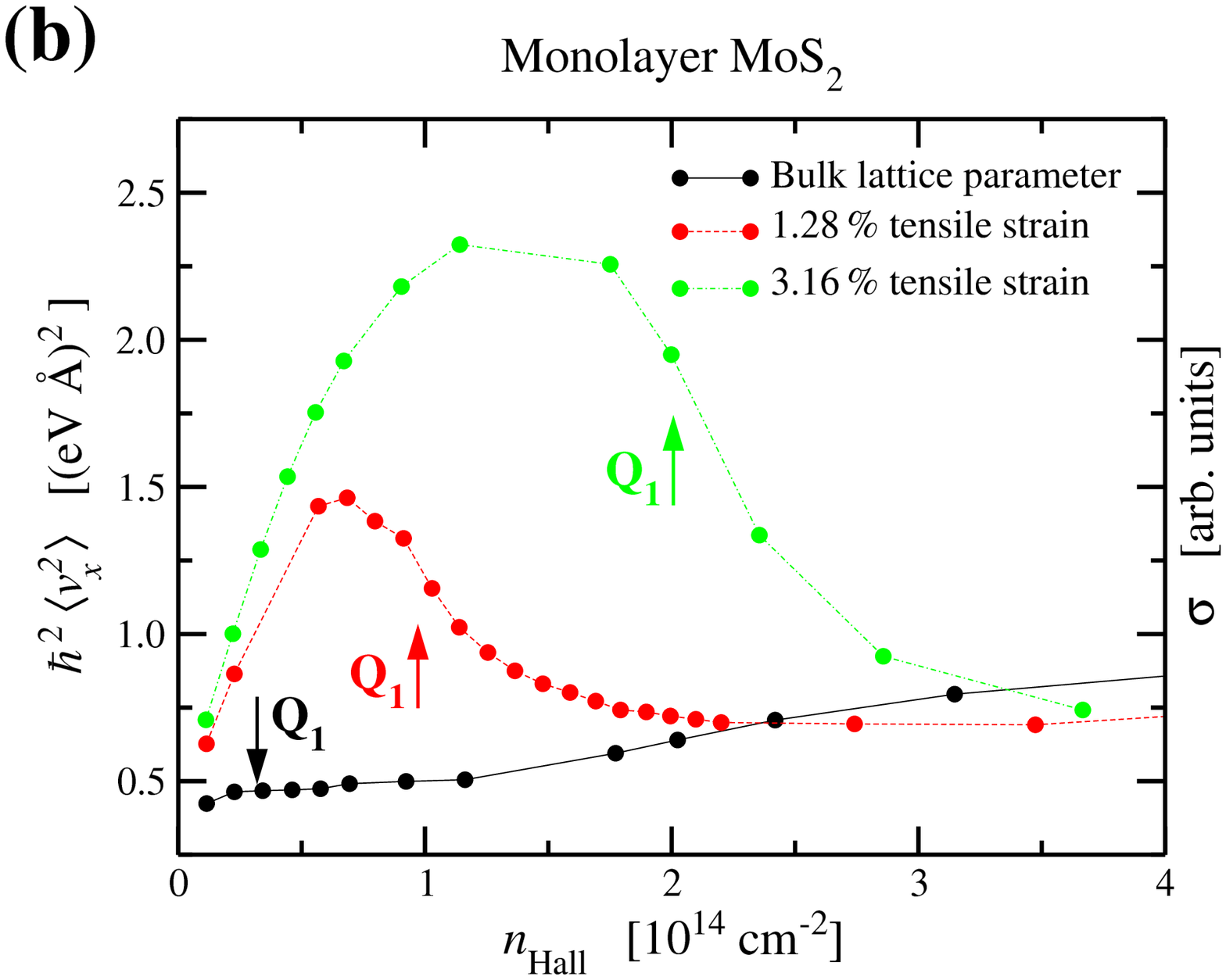}\vspace*{0.5cm}
 \includegraphics[scale=0.4,clip=]{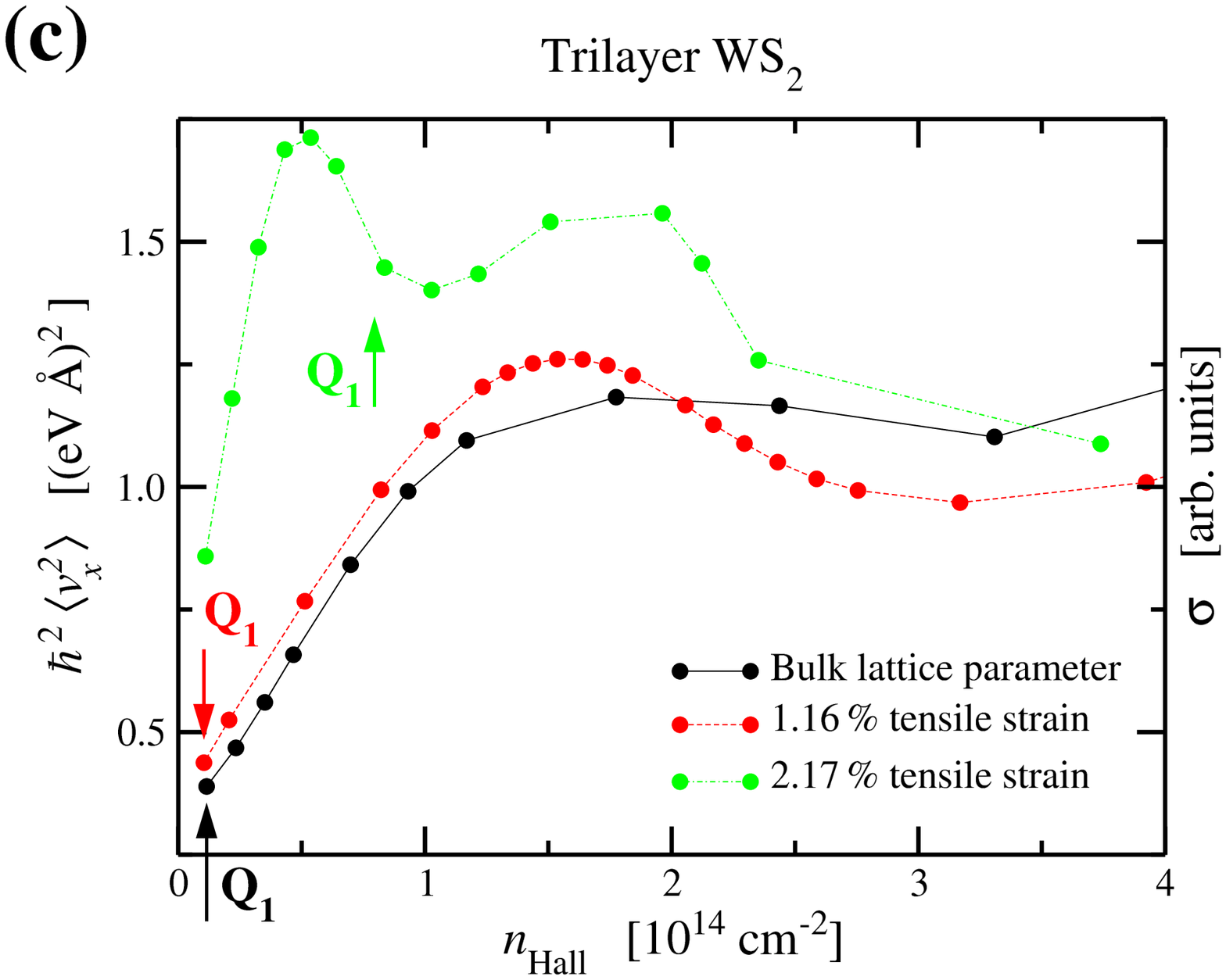}\hspace*{0.5cm}
 \includegraphics[scale=0.4,clip=]{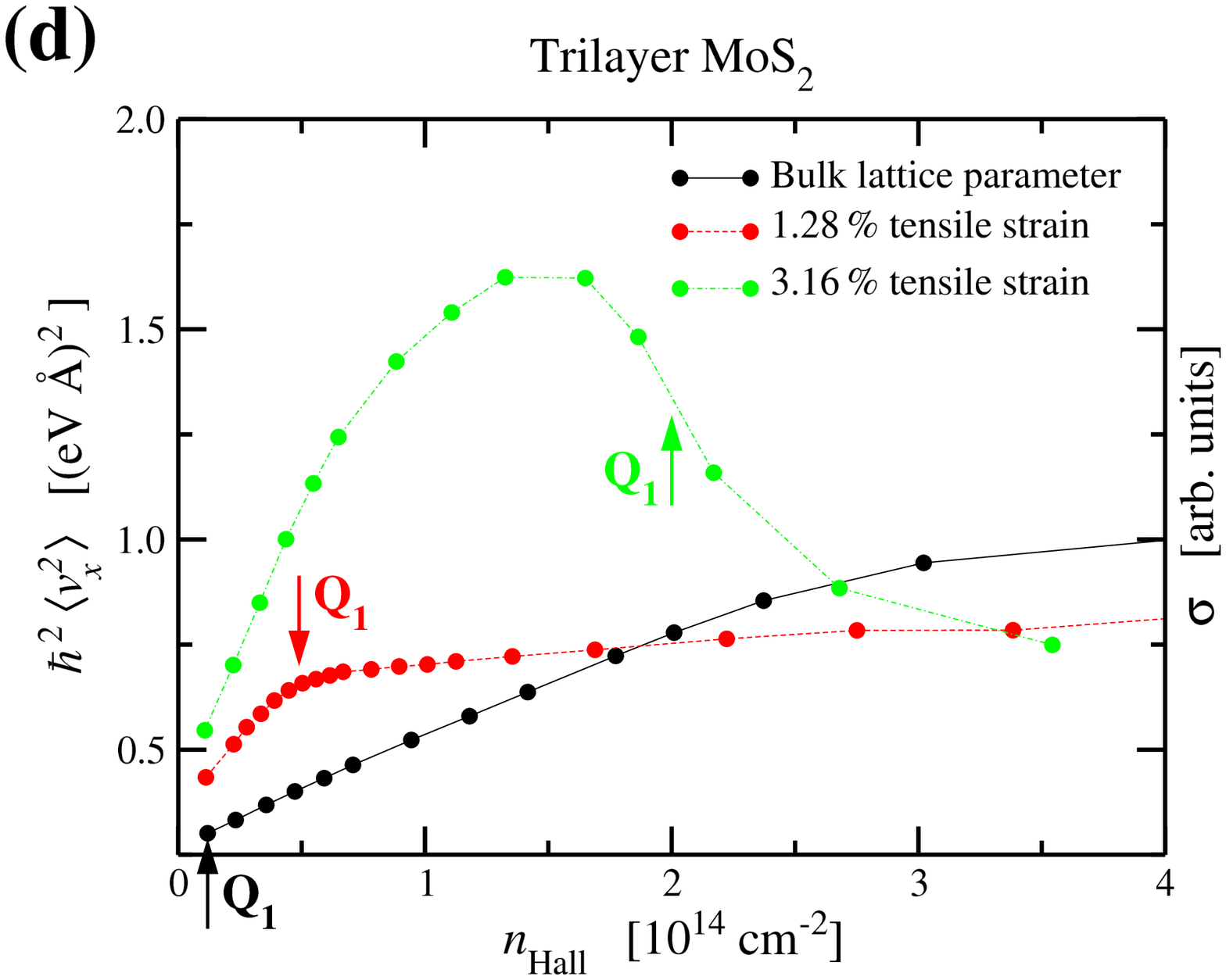}
 \caption{\label{fig:kink_cond} Average squared velocity $\hbar^2\langle v_x^2\rangle$ of (a,c) WS$_2$, (b,d) MoS$_2$ for different number of layers
          with increasing tensile strain $\varepsilon_{xx}=\varepsilon_{yy}\equiv\varepsilon_{||}$ and for $T=300\,\mathrm{K}$.
          The onset of doping of the first band at the Q point (cf. Fig.~\ref{fig:overview}) for $T=0\,\mathrm{K}$ is indicated by small arrows.
          The label ``Bulk lattice parameter'' indicates that the experimental bulk unit cell\cite{podberezskaya2001} was used, i.e. $a=3.160\,\textup{\AA}$
          and $a=3.153\,\textup{\AA}$ for MoS$_2$ and WS$_2$, respectively. The first strain value is actually
          the unit cell as relaxed with PBE+D2 while for the second strain value of MoS$_2$
          we increased unit cell to $a=3.26\,\textup{\AA}$ as found by Jin et al.\cite{jin2015}}
\end{figure*}
The comparison with the simplified model shows that the kink in the conductivity can be used to determine experimentally
the onset of doping of the Q valley in monolayer TMDs and thus the relative position of the conduction-band minimum at Q with respect to the one at K.
In calculations this energy difference between K and Q depends on a lot of parameters (e.g. the level of theory or the unit cell size)
which is why one can also find quite different values in literature. Unfortunately, also experiments were up to now not able to give a clear
answer since for example ARPES only probes the occupied states or light absorption/emission is much stronger for direct transitions. The results
in Figures \ref{fig:velocity} and \ref{fig:compare_model_calc} highlight an easy way to determine the
onset of doping of the Q valley by a conductivity measurement. This is particular important for superconductivity as
the different valleys also exhibit very different electron-phonon coupling\cite{ge2013}.

The average squared velocities shown in Fig.~\ref{fig:velocity} support this picture.
For a quadratic dispersion in 2D the conductivity (which is the squared velocity) is a linearly increasing function of the doping
charge $n$ and has kinks as soon as a new band starts to get doped. The DOS would be constant in this scenario with steps as soon
as a new band enters the bias window. The deviation from this 2D behavior is larger for the trilayer systems in which the kink is
much less pronounced than in the monolayer case. Furthermore,
$\mathrm{DOS}_{E_F}$ has a stronger 2D character in trilayer WS$_2$ than in trilayer MoS$_2$\cite{brumme2015,supmat} which is why the former
has a small decrease in $\hbar^2\langle v_x^2\rangle$ once the second band at Q gets occupied while the latter does not
change at $T=300\,\mathrm{K}$.

A second confirmation that the kink is really related to the onset of doping at Q can be found be applying strain to the system.
It is well-known\cite{peelaers2012,yun2012,shi2013} that strain changes the position of K with respect
to Q considerably. Furthermore, the influence of strain on the conductivity kink is also important from the experimental
point of view as it has been shown that the lattice might relax when going from the bulk to the monolayer limit\cite{jin2015}.
Figure \ref{fig:kink_cond} shows that the maximum in the average squared velocity
(and thus in the conductivity) shifts considerably with applied strain.
In both the monolayer and the trilayer systems the kink can shift by as much as $n_\mathrm{Hall}=1\times10^{14}\,\mathrm{cm}^{-2}$.
Most interestingly as soon as the measured bulk lattice parameter is used no maximum can be observed at all. A clear kink in the conductivity
would thus indicate a relaxation of the lattice as described by Jin et al. in Ref.~\citenum{jin2015}.
Note that if the temperature is decreased the maximum in the conductivity will shift to higher doping values
due to the smaller broadening of $\partial f_{E_F}(T;\varepsilon_{i,\mathbf{k}})/\partial\varepsilon$, cf. Fig.~\ref{fig:velocity}.

In summary, we reported a simple method to extract the scattering time for field-effect doping of TMDs
by comparing the measured Hall mobility $\mu_\mathrm{Hall}^\mathrm{exp}$ with the calculated ratio $\mu_\mathrm{Hall}^\mathrm{theo}/\tau$.
The resulting scattering time can be used to calculate the mean-free path of the charge carriers.
We exemplified the extraction for WS$_2$ for measurements done by Braga et al. in Ref.~\citenum{braga2012}.
For this sample we found that the scattering time for holes is larger than those for electrons and that
the mean-free path for electrons is much shorter. Even if this indicates the breakdown of the semiclassical
BTE for high electron doping of this sample we nevertheless found a good qualitative agreement with experiments.
In the supplemental material\cite{supmat} we provide the data needed to the extract the transport scattering time by a
Hall measurement for monolayers and multilayers of MoS$_2$, MoSe$_2$, MoTe$_2$, WS$_2$, and WSe$_2$.
Finally, we also showed that the onset of doping of the different conduction-band minima at Q can be determined
by a simple conductivity measurement.
As the shape and position of the conductivity peak is influenced by temperature and strain, depending on the
experimental conditions it might be difficult, in some case, to measure a clear peak.
However, a conductance saturation/peak has already been seen for MoS$_2$,\cite{zhang2012} WSe$_2$,\cite{yuan2013,allain2014},
and SnS$_2$\cite{yuan2011} and we thus hope that our calculations stimulate more detailed investigations.

\begin{acknowledgements}
We acknowledge Alberto Morpurgo for giving the hint to have a closer look at the
conductivity vs. doping characteristics.
We furthermore acknowledge financial support of the Graphene Flagship and of the French National ANR funds
within the \textit{Investissements d'Avenir programme} under reference ANR-13-IS10-0003-01.
Computer facilities were provided by PRACE, CINES, CCRT and IDRIS.
\end{acknowledgements}

\bibliography{dichalcogenides_2}

\end{document}